\documentclass[onecolumn,prd,superscriptaddress,nofootinbib,notitlepage]{revtex4-1}
\usepackage{latexsym}
\usepackage{amsmath,amsfonts,graphicx,accents}
\usepackage{amsbsy}
\usepackage{hyperref}
\usepackage{mathrsfs}
\usepackage{color}
\usepackage{psfrag}
\usepackage{enumerate}
\usepackage{amsmath,amssymb,calc,amsfonts}
\usepackage{latexsym}
\usepackage[utf8]{inputenc}
\usepackage[percent]{overpic}
\usepackage[autostyle=false, style=english]{csquotes}
\MakeOuterQuote{"}
\DeclareFontFamily{U}{rsfs}{}         
\DeclareFontShape{U}{rsfs}{m}{n}{<5> rsfs5 <6><7> rsfs7          %
  <8><9><10><10.95><12><14.4><17.28><20.74><24.88> rsfs10}{}     %
\DeclareMathAlphabet{\mathfs}{U}{rsfs}{m}{n}                     %
\definecolor{indiagreen}{rgb}{0.07, 0.53, 0.03}
\def\beq{\begin{eqnarray}}
\def\eeq{\end{eqnarray}}

\def\={\stackrel{\Delta}{=}}

\begin{document}
\title{From Nonextremal to Extremal: Entropy of Reissner-Nordström and Kerr black holes Revisited}

\author{C. Fairoos}\email{fairoos.phy@gmail.com}
\affiliation{T. K. M. College of Arts and Science Kollam-691005, India} 
\author{Chiranjeeb Singha}\email{chiranjeeb.singha@iucaa.in}
\affiliation{Inter-University Centre for Astronomy $\&$ Astrophysics, Post Bag 4, Pune 411 007, India}

\begin{abstract}
In this paper, we derive the entropy of Reissner-Nordström (RN) and Kerr black holes using the Hawking–Gibbons path integral method. We determine the periodicity of the Euclidean time coordinate using two approaches: first, by analyzing the near-horizon geometry, and second, by applying the Chern–Gauss–Bonnet (CGB) theorem. For non-extremal cases, both these methods yield a consistent and unique periodicity, which in turn leads to a well-defined expression for the entropy. In contrast, the extremal case exhibits a crucial difference. The absence of a conical structure in the near-horizon geometry implies that the periodicity of the Euclidean time is no longer uniquely fixed within the Hawking–Gibbons framework. The CGB theorem also fails to constrain the periodicity, as the corresponding Euler characteristic vanishes. As a result, the entropy cannot be uniquely determined using either method.
\end{abstract}

 \maketitle
\section{Introduction}
Black hole thermodynamics has emerged as one of the most profound and far-reaching insights in theoretical physics, revealing a deep and unexpected connection between gravity, quantum theory, and statistical mechanics. Since the groundbreaking work of Bekenstein and Hawking in the 1970s, it has become clear that black holes are not merely classical solutions of general relativity, but also possess thermodynamic properties such as temperature and entropy. Bekenstein's bold proposal that black holes carry an entropy proportional to the area of their event horizon \cite{PhysRevD.7.2333} was later confirmed and refined by Hawking's discovery that black holes radiate thermally due to quantum effects near the horizon, a phenomenon now known as Hawking radiation \cite{Hawking:1975vcx}. This established the Bekenstein–Hawking entropy formula as,
\begin{equation}
S = \frac{k_B A}{4 \ell_{\text{Pl}}^2},
\end{equation}
where $A$ is the horizon area, $k_B$ is Boltzmann’s constant, and $\ell_{\text{Pl}}$ is the Planck length. This result forms a cornerstone in the ongoing effort to unify general relativity and quantum mechanics, as it suggests that the gravitational degrees of freedom possess an underlying microstructure that obeys thermodynamic laws.

Over the past decades, a wide range of theoretical frameworks have been developed to derive, interpret, and generalize black hole entropy. One of the most influential approaches is the Euclidean path integral method, pioneered by Gibbons and Hawking \cite{PhysRevD.15.2752}. In this formalism, the time coordinate is analytically continued to imaginary values, transforming the Lorentzian black hole spacetime into a Riemannian (Euclidean) manifold. The black hole partition function is then approximated semiclassically by evaluating the gravitational action on this Euclidean background. A key insight of this method is that regularity at the event horizon imposes a periodicity condition on the Euclidean time coordinate, which in turn fixes the temperature of the black hole. From the semiclassical relation 
\begin{equation}
S = \beta \frac{\partial I}{\partial \beta} - I,
\end{equation}
where $\beta$ is the inverse temperature, and $I$ is the Euclidean action, one can recover the Bekenstein–Hawking entropy \cite{PhysRevD.47.5400}. For non-extremal black holes, this technique provides a consistent and geometrically transparent derivation of black hole thermodynamics.

However, this picture becomes significantly more subtle in the case of extremal black holes—those whose surface gravity, and hence Hawking temperature, vanishes. Extremal black holes, such as the extremal RN or extremal Kerr black hole solutions, are characterized by horizons that are degenerate in nature, lacking a bifurcate Killing horizon structure. This degeneracy leads to important differences in the topology and geometry of the corresponding Euclidean manifolds. In particular, the Euclidean continuation of an extremal black hole does not exhibit the smooth "cigar"-like geometry found in non-extremal cases, which casts doubt on the applicability of the standard Euclidean techniques. Some analyses, including the work of Das et al. \cite{Das:1996rn}, argue that extremal black holes have zero entropy, emphasizing the discontinuity between extremal and non-extremal geometries and the absence of a regular Euclidean section. This view challenges the notion that extremal black holes are the smooth limit of non-extremal ones, suggesting instead that they may belong to a qualitatively different class of gravitational objects \cite{Hawking:1994ii,Hawking:1995fd}.

On the other hand, insights from string theory have provided compelling evidence in favor of a finite, non-zero entropy for certain extremal black holes. In a landmark calculation, Strominger and Vafa \cite{Strominger:1996sh} demonstrated that the entropy of a class of supersymmetric extremal black holes in five dimensions could be exactly reproduced by counting the degeneracy of bound D-brane states in the weak-coupling regime. This result provided a statistical-mechanical explanation of black hole entropy and confirmed the validity of the Bekenstein–Hawking formula from a microscopic standpoint. Subsequent works, such as those by Mathur \cite{Mathur:2005zp}, have further developed this perspective, proposing that black hole microstates correspond to smooth, horizonless geometries, a viewpoint encapsulated in the fuzzball conjecture. These findings indicate that extremal black holes may indeed possess a finite entropy, but that their origin and interpretation are sensitive to the theoretical context, especially in frameworks beyond semiclassical general relativity \cite{Bardeen:1999px, Sen:2007qy, Hawking:1994ii, Marolf:2008tx,Ghosh:1996gp, Cai:1997ih}. This ongoing tension underscores the need to reconcile different approaches to black hole entropy, particularly in the extremal regime.

A complementary line of investigation emphasizes the role of topology in black hole thermodynamics. In this context, the CGB theorem has proven to be a powerful tool for understanding the geometric and topological origin of entropy \cite{Banados:1993qp, Gibbons:1979xm, Gibbons:1994ff, PhysRevD.51.4315,Liberati:1997sp}. In higher-dimensional gravity theories or theories with higher-curvature corrections, the Wald entropy formula generalizes the area law, and often the entropy can be expressed in terms of topological invariants. The work of Bañados, Teitelboim, and Zanelli \cite{Banados:1993qp}, along with that of Wu \cite{Wu:2000pp}, has shown that the entropy of certain black hole solutions can be directly linked to the Euler characteristic of their Euclidean sections. More recently, a novel result by Hughes \cite{Hughes:2025ved} has established a connection between the Hawking temperature and the Euler characteristic in the context of Wick-rotated Schwarzschild and de Sitter geometries, further reinforcing the idea that black hole entropy is fundamentally topological in nature.

However, the topological approach, too, faces challenges when extended to extremal black holes. The application of the CGB theorem relies on the structure of the Euclidean manifold, and for extremal black holes, the Euclidean sections often lack the required topological features, such as a conical defect or a non-trivial Euler characteristic. This has led to ambiguities and debates regarding whether extremal black holes can be treated as limiting cases of non-extremal geometries, or whether they must be regarded as qualitatively distinct. The breakdown of the path integral and topological methods in the extremal limit reflects a broader issue: the tools of semiclassical gravity, which work well for non-extremal spacetimes, may need to be fundamentally rethought or extended in order to properly capture the physics of extremal configurations.\\

In light of these considerations, the present work revisits the calculation of black hole entropy through the gravitational path integral, aiming to shed light on both non-extremal and extremal scenarios within a unified framework. We begin with a detailed exposition of the Gibbons–Hawking approach, using the Schwarzschild black hole as a pedagogical example to illustrate how thermodynamic quantities naturally arise from the Euclidean formalism. We then generalize this framework to near-extremal RN and Kerr black holes, exploring how the inclusion of higher-curvature terms and topological invariants, particularly via the CGB theorem, modifies the entropy calculation. Finally, we turn our attention to the extremal case, critically examining the limitations of the Euclidean path integral and the difficulties in applying topological methods. Through this analysis, we aim to clarify the conceptual status of extremal black hole entropy and to contribute to the broader understanding of black holes as thermodynamic and quantum objects.\\

This paper is organized as follows. In section \ref{pathintegral}, we review the Hawking–Gibbons path integral approach to black hole thermodynamics, highlighting its foundational role in understanding black hole entropy. In section \ref{Non-extremal}, we analyze the entropy of non-extremal black holes, focusing on the RN and Kerr solutions. Section~\ref{extremal} is devoted to the study of extremal black holes, where we examine the entropy of the extremal RN and Kerr spacetimes and discuss the subtleties that arise in the extremal limit. Finally, in section \ref{dis}, we summarize our findings and conclude the paper with a discussion of potential implications.\\

\emph{Notations and conventions:}  
Throughout this paper, we use the metric signature $(-,+,+,+)$, corresponding to the Minkowski metric in $3+1$ dimensions expressed in Cartesian coordinates as $\mathrm{diag.}(-1, +1, +1, +1)$. We also work in natural units by setting $G = c =\hbar=1$.

\section{Hawking-Gibbons Path Integral Approach to Black Hole Thermodynamics}\label{pathintegral}

The Hawking-Gibbons path integral method provides a semi-classical approach to understanding black hole thermodynamics by evaluating the gravitational partition function in Euclidean quantum gravity. The key idea is to analytically continue the spacetime metric to imaginary (Euclidean) time and compute the path integral over Euclidean geometries\cite{PhysRevD.15.2752}. That can be computed following equation
\begin{equation}\label{partition}
Z = \int \mathcal{D}[g]\, e^{-I_E[g]}~,
\end{equation}
where $I_E$ is the Euclidean gravitational action, and $Z$ serves as the partition function. The leading contribution comes from classical solutions (saddle points) of the Euclidean Einstein equations. Let us briefly outline how the thermodynamic quantities of the Schwarzschild black hole are obtained using this prescription. As we plan to extend our discussion to other spacetime geometries, we start with a general spherically symmetric stationary metric in the Lorentzian signature as,
\begin{equation}\label{metric}
ds^2 = -f(r)dt^2 + \frac{dr^2}{f(r)} + r^2 d\Omega^2~.
\end{equation}
For the Schwarzschild case, we have the metric function $f(r)=1 - \frac{2M}{r}$. The corresponding Euclidean metric is obtained by performing the Wick Rotation, i.e.,  \(t \to -i\tau\). Then the above metric looks like,
\begin{equation}
ds^2_E = \left(1 - \frac{2M}{r}\right) d\tau^2 + \left(1 - \frac{2M}{r}\right)^{-1} dr^2 + r^2 d\Omega^2~.
\end{equation}
 This change is crucial for formulating a quantum theory of gravity in thermal equilibrium, as it allows us to define a well-behaved partition function in terms of the Euclidean path integral. After Wick rotation, the Schwarzschild metric becomes positive-definite, and the region near the event horizon at \( r = 2M \) acquires a geometry similar to a two-dimensional plane in polar coordinates. To prevent a conical singularity at the horizon (analogous to a sharp point in a cone), the Euclidean time coordinate \( \tau \) must be periodic with a specific period \( \beta = 8\pi M \). This periodicity is interpreted as the inverse temperature of the black hole, linking the geometry to thermodynamic properties. The resulting Euclidean manifold has the shape of a ``cigar'' — smooth at the tip (the horizon) and asymptotically cylindrical. This regularity condition is essential for the Gibbons-Hawking method, as it ensures the action is finite and the path integral is well-defined. Without this regularization, the geometry would be singular, and the thermal interpretation of the path integral would break down. To use Eq. \ref{partition}, we consider the Euclidean action which contains the Einstein-Hilbert term and the Gibbons-Hawking-York boundary term:
\begin{equation}\label{Scw_action}
I_E = -\frac{1}{16\pi} \int_{\mathcal{M}} R \sqrt{g} \, d^4x - \frac{1}{8\pi} \int_{\partial\mathcal{M}} (K - K_0) \sqrt{h} \, d^3x~.
\end{equation}
Here, \( R \) is the Ricci scalar of the Euclidean Schwarzschild manifold, which vanishes (\( R = 0 \)) for the Schwarzschild solution in vacuum. Thus, only the boundary term contributes to the Euclidean action. The extrinsic curvature \( K \) is computed on the boundary \( \partial\mathcal{M} \), which is taken at large radial coordinate \( r \to \infty \). However, the pure Schwarzschild contribution diverges at infinity. To obtain a finite and physically meaningful result, we subtract the corresponding boundary term for flat Euclidean space, whose extrinsic curvature is denoted by \( K_0 \). This subtraction ensures that the action is normalized with respect to flat space, effectively setting the zero-point of energy such that Minkowski spacetime has zero action. This regularization is essential to correctly compute the thermodynamic quantities, such as the free energy and entropy of the black hole, using the semiclassical approximation to the path integral. The above integration results,
\begin{equation}
I_E = \frac{\beta M}{2}~.
\end{equation}
Now, the entropy \(S\) is derived from the thermodynamic relation:
\begin{equation}
S = \beta \frac{\partial I_E}{\partial \beta} - I_E= 4\pi M^2 =  \frac{A}{4}~.
\end{equation}
This is the Bekenstein-Hawking entropy. In summary, the requirement of regularity in Euclidean time determines $\beta$, and the evaluation of the Euclidean action leads to the entropy. This method elegantly connects gravitational dynamics with thermal properties and is foundational in quantum gravity studies \cite{Almheiri2020, Almheiri2020_2, Penington2022}.\\

Interestingly, an alternative method exists to calculate the periodicity of the Euclidean time coordinate due to the CGB theorem. According to this, the topological Euler characteristics $\chi$ of the manifold $\mathcal{M}$ is defined by \cite{Chern1944, Chern1945,Mielke2017},
\begin{equation}
\chi(\mathcal{M})=\frac{1}{32 \pi^2}\int_{\mathcal{M}}\sqrt{-g}d^4x\left(R^2-4R_{\mu\nu}R^{\mu\nu}+R_{\mu \nu \rho \sigma}R^{\mu \nu \rho \sigma}\right)~.
\end{equation}
Note that the validity of the theorem requires the manifold to be compact, and therefore, we cannot directly use the definition for most of the vacuum spacetimes, such as Schwarzschild. However, the theorem can be applied in the Wick rotated Euclidean space, where our attention is restricted to the near-horizon regime, as pointed out in \cite {Hughes:2025ved}. Besides, for the Schwarzschild case, the first two terms in the integral are zero, and the Kretschmann scalar $R_{\mu \nu \rho \sigma}R^{\mu \nu \rho \sigma}$ falls off sufficiently fast, making no contribution from the asymptotic region to $\chi(\mathcal{M})$. The expression for the Euler characteristic for the Euclidean Schwarzschild spacetime is,
\begin{equation*}
    \chi(\mathcal{M}_E) = \frac{1}{32 \pi^2}\int_0^\beta d\tau \int_{2M}^\infty dr \int_\Sigma \sqrt{g}R_{\mu \nu \rho \sigma}R^{\mu \nu \rho \sigma}~.
\end{equation*}
A direct calculation yields, 
\begin{equation*}
    R_{\mu \nu \rho \sigma}R^{\mu \nu \rho \sigma}=\frac{48 M^2}{r^6}~.
\end{equation*}
Now, the period,
\begin{equation*}
    \beta = \chi(\mathcal{M}_E)4\pi M~.
\end{equation*}
As the topology of the spacetime is $D^2 \times \mathbb{S}^2$, we have,
\begin{equation*}
    \chi(\mathcal{M}_E) = \chi(D^2)\times \chi(\mathbb{S}^2)= 1\times2=2~.
\end{equation*}
Substituting for $\chi(\mathcal{m}_E)$ into the expression for $\beta$ gives the expected result, i.e.,
\begin{equation*}
    \beta = 8\pi M~.
\end{equation*}
As $\beta$ is understood as the inverse Hawking temperature, the above description connects the Hawking temperature to the topology of the Euclidean geometry.\\

In this section, we have outlined the derivation of black hole entropy using the Hawking-Gibbons path integral approach for the Schwarzschild black hole. Also, we have explored two equivalent methods to determine $\beta$. We now turn to more general stationary black hole solutions—specifically, the non-extremal RN and Kerr spacetimes. These solutions introduce electric charge and angular momentum, respectively, enriching the thermodynamic structure and geometry of black holes. Studying the Euclidean path integral in these settings not only provides a deeper understanding of black hole entropy and thermodynamics but also sets the stage for addressing the more subtle case of extremal black holes. Since the extremal limit leads to qualitatively different geometric and topological features, it is essential to first analyze the thermodynamics of their non-extremal counterparts within the same formalism. This will help us understand the modifications and challenges involved in extending the approach to extremal RN and Kerr black holes.

\section{Entropy of Non-extremal Reissner-Nordström and Kerr Black Holes}\label{Non-extremal}

The Euclidean gravitational action for RN spacetime can be obtained by modifying Eq. \ref{Scw_action} by including an electromagnetic term,
\begin{equation}
 I_E = -\frac{1}{16\pi} \int_{\mathcal{M}} \left(R-F_{\mu\nu}F^{\mu\nu}\right) \sqrt{g} \, d^4x - \frac{1}{8\pi} \int_{\partial \mathcal{M}} \left(K-K_0\right) \sqrt{h} \, d^3x.
 \end{equation}
Here $F_{\mu\nu}$ is the electromagnetic field tensor. As the Einstein equations hold in the bulk, only the boundary term contributes from the gravitational sector. However, the electromagnetic Lagrangian has the following contribution,
\begin{eqnarray*}
\frac{1}{16\pi}\int_{\mathcal{M}}d^4x\sqrt{g}F_{\mu\nu}F^{\mu\nu} &=& \frac{\beta}{4}\int_{r_+}^{r_b}r^2 dr \left(\frac{2Q^2}{r^4}\right) \\
&=& -\frac{\beta}{2}Q^2\left(\frac{1}{r_b}-\frac{1}{r_+}\right),
\end{eqnarray*}
where the integration over periodic time coordinate yields $\beta$. To arrive the above result, we have used $F_{\mu\nu}F^{\mu\nu} = \frac{2Q^2}{r^4}$. The radial integration is from the horizon ($r_+=M+\sqrt{M^2-Q^2}$) to some constant $r$ surface, i.e., $r=r_b$. Note that the final contribution from the electromagnetic Lagrangian is obtained by taking the limit $r_b \to \infty$. To calculate the boundary contribution to the action, we consider the induced 3-metric on the boundary of the Euclideanized version of Eq. \ref{metric},
\[ ds^2_{\partial \mathcal{M}} = f(r_b) d\tau^2 + r_b^2 d\Omega^2. \]

The unit normal to the boundary (which is a constant-r surface) is:
\[ n^\mu = \left(0, \sqrt{f(r)}, 0, 0\right) \quad \Rightarrow \quad n_\mu = \left(0, \frac{1}{\sqrt{f(r)}}, 0, 0\right) .\]

The trace of extrinsic curvature is obtained as,
\[ K = h^{ab} \nabla_a n_b = \nabla_\mu n^\mu = \frac{1}{\sqrt{g}} \partial_\mu \left(\sqrt{g} n^\mu\right). \]

As the determinant \(\sqrt{g} = r^2\sin\theta \), the trace of extrinsic curvature can be expressed in terms of the metric function $f(r)$ as,
\[ K = \frac{1}{\sqrt{g}} \partial_r (\sqrt{g} n^r) = \frac{1}{r^2 \sin\theta} \partial_r \left( r^2 \sin\theta \sqrt{f(r)} \right) = \frac{2 \sqrt{f(r)}}{r} + \frac{f'(r)}{2 \sqrt{f(r)}}. \]
The extrinsic curvature on the flat boundary is constructed from $n^r=(0,1,0,0)$, giving $K_0=\frac{2}{r}$. Therefore, the boundary term becomes,
 \begin{eqnarray}
 -\frac{1}{8\pi} \int_{\partial \mathcal{M}} \left(K-K_0\right) \sqrt{h} \, d^3x 
 &=& -\frac{1}{8\pi} \int_0^\beta d\tau \int_0^\pi d\theta \int_0^{2\pi} d\phi \,  \left(K-K_0\right) r^2 \sin\theta \sqrt{f(r)} \\ \nonumber
 &=& -\frac{\beta}{2} r^2 \left( \frac{2 f(r)}{r} + \frac{f'(r)}{2} -\frac{2}{r} \sqrt{f(r)}\right)~. 
\end{eqnarray}
  Now, the total action becomes,
\begin{equation}\label{Eucl_action_gene}
I_E = -\frac{\beta}{2} \left(-\frac{Q^2}{r_+}+ 2r f(r) + \frac{r^2 f'(r)}{2} -2 r \sqrt{f(r)}+ \frac{Q^2}{r}  \right)~. 
\end{equation}
Note that all these terms are evaluated at $r=r_b$. In the end, we will push $r_b \to \infty$. The Euclidean action obtained in Eq. \ref{Eucl_action_gene} is valid for any stationary spherically symmetric spacetimes. Substituting $f(r) = 1-\frac{2M}{r}$ and $Q=0$ gives the expression for the Euclidean action for Schwarzschild spacetime mentioned in the previous section. For the Reissner–Nordström black hole, the metric function is:
\begin{equation}
f(r) = 1 - \frac{2M}{r} + \frac{Q^2}{r^2}.
\end{equation}
Eq. \ref{Eucl_action_gene} now becomes,
\begin{equation}
I_E =- \frac{\beta}{2} \left[ -\frac{Q^2}{r_+} + 2r_b \left(1 - \frac{2M}{r_b} + \frac{Q^2}{r_b^2} \right) + \frac{r_b^2}{2}\left(\frac{2M}{r_b^2}-\frac{2Q^2}{r_b^3}\right)-2r_b\sqrt{f(r_b)}+\frac{Q^2}{r_b} \right],
\end{equation}
In the limit $r_b \to \infty$, we have,
\begin{equation}
I_E= \frac{\beta}{2} \left(M+\frac{Q^2}{r_+}\right),
\end{equation}
The entropy is given by,
\begin{equation}\label{RN_entropy}
S = \beta \left( \frac{\partial I_E}{\partial \beta} \right) - I_E =\frac{\beta}{2}\sqrt{M^2-Q^2}.
\end{equation}
It remains to calculate the periodicity of the Euclidean time coordinate near the horizon. To this extent, we expand \( f(r) \) about \( r = r_+ \) and express the metric using a new radial coordinate $\rho$, where $\rho = 2\sqrt{\frac{r - r_+}{f'(r_+)}}$. Near the horizon, the metric looks,
\begin{equation}
ds^2 \approx \left( \frac{f'(r_+)^2}{4} \rho^2 \right) d\tau^2 + d\rho^2 + r_+^2 d\Omega^2.
\end{equation}
The first two terms describe a flat 2D space in polar coordinates, i.e., in the form of $dR^2+R^2d\theta^2$ with $\theta = \theta +2\pi$. To avoid a conical singularity at \( \rho = 0 \), the angular coordinate \( \tau \) must satisfy the periodic condition $\tau = \tau +4\pi/f'(r_+)$. Therefore, the periodicity,
\begin{equation}
\beta = \frac{4\pi}{f'(r_+)} = \frac{2\pi r_+^2}{\sqrt{M^2-Q^2}},
\end{equation}
Before proceeding to calculate the entropy, our primary objective, we first verify the applicability of the CGB theorem for the Euclidean RN spacetime. To this end, the expression for the Euler characteristic takes the form,
\begin{equation}\label{chi_RN}
    \chi(\mathcal{M}_E) = \frac{1}{32 \pi^2}\int_0^\beta d\tau \int_{r_+}^\infty dr \int_\Sigma \sqrt{g} d^2x\left(R_{\mu \nu \rho \sigma}R^{\mu \nu \rho \sigma}-4 R_{\mu \nu}R^{\mu \nu}\right).
\end{equation}
It is important to emphasize that, unlike in the Schwarzschild case, the Ricci tensor does not vanish due to the presence of electric charge in the RN black hole solution. Moreover, it can be explicitly verified that the integrands decay sufficiently fast at spatial infinity, ensuring that the asymptotic contributions from both curvature terms vanish. Consequently, the application of the CGB theorem is justified for evaluating the inverse temperature $\beta$. A direct computation yields the following curvature invariants:
\begin{eqnarray*}
    R_{\mu \nu \rho \sigma}R^{\mu \nu \rho \sigma}&=&\frac{8}{r^8}\left(6M^2r^2-12 M Q^2 r+7 Q^4\right)~,\\
    R_{\mu \nu}R^{\mu \nu}&=& \frac{4 Q^2}{r^8}~.
\end{eqnarray*}
Substituting these expressions into the CGB integral, the resulting Euler characteristic is found to be,
\begin{eqnarray*}
    \chi(\mathcal{M}_E) = \frac{\beta \sqrt{M^2-Q^2}}{\pi Q^4}\left(M-\sqrt{M^2-Q^2}\right)^2~.
\end{eqnarray*}
Imposing the topological requirement $ \chi(\mathcal{M}_E) =2$ and simplifying the resulting expression, we obtain the inverse temperature:
\begin{equation*}
    \beta=\frac{2\pi}{\sqrt{M^2-Q^2}}r_+^2.
\end{equation*}
Thus, we have once again successfully derived the expression for $\beta$ purely from the topological properties of the Euclideanized spacetime, confirming the robustness of the CGB approach in this context.
Now, Eq. \ref{RN_entropy} can be expressed in the familiar Bekenstein-Hawking entropy formula, i.e., 
\begin{equation}
S = \pi r_+^2.
\end{equation}
We now focus on the Kerr black hole solution. The Kerr metric represents the gravitational field outside a rotating, massive object that is both stationary and axisymmetric. It is also regarded as the ultimate configuration reached by a collapsing star, uniquely specified by mass and angular momentum. Unlike Schwarzschild or RN black holes, the Kerr black hole exhibits significantly different thermodynamic behavior due to its more complex causal structure. As a result, it serves as a crucial model for investigating the physical characteristics of astrophysical systems and for testing theoretical predictions concerning black hole thermodynamics. The entropy of a non-extremal Kerr black hole can also be obtained using the same approach as detailed above. The metric in Boyer–Lindquist coordinates is given by \cite{Oneill},
\begin{equation*}
ds^2=-\frac{\rho^2 \Delta}{\Xi}dt^2 + \frac{\rho^2}{\Delta}dr^2+\rho^2 d\theta^2 + \frac{\Xi}{\rho^2}\sin^2\theta\left(d\phi-\frac{2Ma r}{\Xi}dt\right)^2,
\end{equation*}
where,
\begin{eqnarray*}
\rho^2=r^2+a^2\cos^2\theta&;& \quad \Delta=r^2-2Mr+a^2;\\
\Xi &=& \left(r^2+a^2\right)^2-a^2\Delta \sin^2\theta.
\end{eqnarray*}
The Euclidean version of the metric obtained after the Wick rotation yields \cite{Brown:1990fk},
\begin{equation}\label{Kerr_Eu}
ds^2=\frac{\rho^2 \Delta}{\Xi}d\tau^2 + \frac{\rho^2}{\Delta}dr^2+\rho^2 d\theta^2 - \frac{\Xi}{\rho^2}\sin^2\theta\left(d\phi-\frac{2Ma r}{\Xi}d\tau\right)^2,
\end{equation}
when computing the on-shell action for the Kerr solution, there is no bulk contribution, as the geometry is Ricci flat. All the information is in the Gibbons-Hawking term evaluated on the boundary cut-off surface, i.e.,
\begin{equation}
I_E = -\frac{1}{8\pi}\int_0^\beta d\tau \ \int_0^\pi d\theta \ \int_0^{2\pi} d\phi \sqrt{h}\left(K-K_0\right),
\end{equation}
where $K$ is the trace of extrinsic curvature on the boundary and $K_0$ is the trace of extrinsic curvature on the same boundary but with $M=a=0$.
\begin{equation}
K = \frac{1}{2\sqrt{\Delta} \rho^3}\frac{d}{dr}\left(\rho^2 \Delta\right); \quad \text{and}\quad \sqrt{h}=\sin\theta \sqrt{\Delta}\rho~.
\end{equation}
One can find,
\begin{equation*}
K = \frac{1}{\sqrt{\Delta} \rho^3}\left(\rho^2(r-M)+r \Delta\right); \quad \text{and}\quad K_0=\frac{2}{r}.
\end{equation*}
The Euclidean action becomes,
\begin{eqnarray*}
I_E &=& -\frac{\beta}{4}\int_0^\pi d\theta \sin\theta \sqrt{\Delta}\rho \Bigg[\frac{1}{\sqrt{\Delta} \rho^3}\left(\rho^2(r-M)+r\Delta\right)-\frac{2}{r}\Bigg]\\
&=&-\frac{\beta}{4}\int_0^\pi \left((r-M)+\frac{r\Delta}{\rho^2}-\frac{2\rho \sqrt{\Delta}}{r}\right)\sin\theta d\theta.
\end{eqnarray*}
This integration can be evaluated to yield,
\begin{eqnarray*}
I_E &=& -\frac{\beta}{4}\Bigg[ 2(r-M)+ \frac{2\Delta}{a}\tan^{-1}\frac{a}{r}-\frac{2\sqrt{\Delta}}{r}\sqrt{r^2+a^2}-\frac{\sqrt{\Delta}r}{a}\ln \left(\frac{a+\sqrt{r^2+a^2}}{-a+\sqrt{r^2+a^2}}\right)\Bigg].
\end{eqnarray*}
Now, this expression should be evaluated in the limit $r \to \infty$, giving,
\begin{equation}
I_E = \frac{\beta M}{2}.
\end{equation}
The entropy is obtained from the standard relation, 
\begin{equation*}
S = \left(\beta \frac{\partial}{\partial \beta}-1\right)I_E = \frac{1}{2}\beta^2 \frac{\partial M}{\partial \beta}.
\end{equation*}
To obtain $\beta$, we will consider two methods as before, i.e., inspecting the near-horizon Kerr metric and the CGB theorem. To obtain the near-horizon geometry, we consider the coordinate change $r=r_++\epsilon$. The quantities $\rho, \Delta$, and $\Xi$ become,
\begin{eqnarray*}
\rho^2&=&r_+^2+a^2\cos^2\theta \equiv \rho_+^2; \quad \Xi = \left(r_+^2+a^2\right)^2 \equiv \Xi_+;\\
\Delta &=& 2(r_+-M)\epsilon \equiv \Delta_+.
\end{eqnarray*}
The near-horizon Kerr metric now becomes,
\begin{equation}
ds^2=\frac{\rho_+^2 \Delta_+}{\Xi_+} d\tau^2 + \frac{r_+^2+a^2\cos^2\theta}{2(r_+-M)\epsilon} d\epsilon^2 +\rho_+^2 d\theta^2 - \frac{\Xi_+}{\rho_+^2}\sin^2\theta\left(d\phi-\frac{a }{(r_+^2+a^2)}d\tau\right)^2,
\end{equation}
Now, consider the coordinate transformation,
\begin{equation*}
R = \left(\frac{2(r_+^2+a^2\cos^2\theta)}{r_+-M}\right)^{\frac{1}{2}}\sqrt{\epsilon}.
\end{equation*}
Then the metric takes the following form,
\begin{equation}
ds^2=\frac{\left(r_+-M\right)^2}{\left(r_+^2+a^2\right)^2}R^2 d\tau^2+ dR^2+ \rho_+^2 d\theta^2 - \frac{\Xi_+}{\rho_+^2}\sin^2\theta\left(d\phi-\frac{a}{r_+^2+a^2}d\tau\right)^2.
\end{equation}
Comparing with $R^2d\theta'^2+dR^2$,  where $\theta'$ is some angular coordinate with periodicity $2\pi$, one can obtain the periodicity ($\beta$) for the Euclidean time coordinate, i.e.,
\begin{equation}\label{beta_Kerr}
\beta = 2\pi \frac{r_+^2+a^2}{r_+-M}.
\end{equation}
Now, following the CGB theorem, the expression for the Euler characteristic is given by,
\begin{equation*}
    \chi(\mathcal{M}_E) = \frac{1}{32 \pi^2}\int_0^\beta d\tau \int_{r_+}^\infty dr \int_\Sigma \sqrt{g} d^2x R_{\mu \nu \rho \sigma}R^{\mu \nu \rho \sigma}.
\end{equation*}
Here, $r_+$ is the Kerr horizon. The Kretschmann scalar is calculated from Eq. \ref{Kerr_Eu}, gives \cite{Visser:2007fj},
\begin{equation*}
    R_{\mu \nu \rho \sigma}R^{\mu \nu \rho \sigma}=\frac{48 M^2\left(r^2-a^2\cos^2\theta\right)\Big[\left(r^2+a^2\cos^2\theta\right)^2-16r^2a^2\cos^2\theta\Big]}{\left(r^2+a^2\cos^2\theta\right)^6}.
\end{equation*}
Integrating the above expression yields,
\begin{eqnarray*}
    \chi(\mathcal{M}_E) &=&\frac{\beta}{2\pi a^2 M}\Big[a^2+M\left(\sqrt{M^2-a^2}-M\right)\Big]\\
    &=& \frac{\beta}{\pi}\frac{\left(r_+-M\right)}{\left(r_+^2+a^2\right)}.
\end{eqnarray*}
Since $ \chi(\mathcal{M}_E)=2$, the inverse temperature becomes,
\begin{equation*}
    \beta=2\pi \frac{r_+^2+a^2}{r_+-M}.
\end{equation*}
This result is in exact agreement with Eq. \ref{beta_Kerr}, thereby confirming the validity of the CGB theorem for Euclidean Kerr spacetime. As in the cases of Schwarzschild and RN spacetimes, the CGB theorem once again yields the correct expression for the inverse temperature $\beta$. 
With this result in hand, the entropy can be computed using the standard thermodynamic relation:
\begin{equation}
S= \frac{1}{2}\beta^2 \frac{\partial M}{\partial \beta} = \pi\left(r_+^2+a^2\right).
\end{equation}
In this section, we have explicitly calculated the entropy of both RN and Kerr black holes using the Hawking-Gibbons path integral approach. In both cases, the expression for entropy is one-fourth of the horizon area. In the following section, we consider the same method for calculating the entropy of extremal black holes.

\section{Entropy of Extremal Black Holes}\label{extremal}
For the extremal RN Black Hole, we have the condition $M=Q$, and the Wicks' rotated metric takes the form
\begin{equation}
ds^2 = \bigg(1-\frac{M}{r}\bigg)^2 d\tau^2+\frac{dr^2}{\bigg(1-\frac{M}{r}\bigg)^2}+r^2d\Omega^2.
\end{equation}
The classical Euclidean action for the above metric can be calculated using Eq. \ref{Eucl_action_gene} followed by taking the asymptotic limit on the radial coordinate. One obtains,
\begin{equation}\label{extre_action}
I_E=\frac{\beta M}{2}.
\end{equation}
To obtain an expression for $\beta$ in terms of the mass of the black hole, we should analyse the near-horizon geometry. To this extent, let's consider the following coordinate transformation: $r=M+\rho$ where $\rho<<M$. The metric now takes the following form,
\begin{equation}
ds^2=\frac{\rho^2}{M^2}d\tau^2 + \frac{M^2}{\rho^2} d\rho^2.
\end{equation}
This metric, unlike for the non-extremal case, can not be expressed in the form of $dR^2+R^2d\theta^2$, and therefore, there is no "tip" of a cone — i.e., no origin where the geometry becomes singular unless the time coordinate has a specific periodicity. This means that the Euclidean solution can be identified with any period $\beta$. Thus, we cannot obtain a unique expression for the entropy solely based on this method. Naturally, we would like to apply the CGB theorem to the extremal RN black hole to see whether a similar conclusion holds. To this end, we evaluate Eq. \ref{chi_RN} using the Euclidean metric for the extremal RN black hole. One obtains,
\begin{equation}
     \chi(\mathcal{M}_E) = \frac{\beta}{ \pi} \int_{M}^\infty dr \frac{M^2}{r^6}\left(5 M^2-12 M r +6 r^2\right) =0.
\end{equation}
As the near-horizon Euclidean metric in this case has the topology of $\mathbb{R} \times \mathbb{S}^1 \times \mathbb{S}^2$, the Euler characteristic vanishes, rendering the periodicity $\beta$ undetermined. This observation is consistent with the result from the Hawking-Gibbons path integral approach. In essence, the topology of the near-horizon geometry, along with the application of the CGB theorem, fails to uniquely fix the value of $\beta$, thereby making the entropy ill-defined.\\

A similar conclusion holds for the extremal Kerr black hole. While it is possible to construct a well-defined Euclidean action for the extremal Kerr spacetime, the absence of a conical structure in the near-horizon geometry means there is no preferred periodicity $\beta$. This feature is reflected in the CGB theorem as well: the CGB integral vanishes identically, just as in the extremal RN case. Consequently, the failure to determine a unique $\beta$ leads to the breakdown of entropy calculation using the Hawking-Gibbons path integral method for extremal black holes. This reinforces the conclusion through both geometric and topological arguments provided by the CGB theorem.\\

\section{discussion}\label{dis}

In this paper, we have undertaken a detailed derivation of the entropy associated with RN and Kerr black holes within the framework of the Euclidean path integral approach developed by Hawking and Gibbons. Our analysis began with examining non-extremal black holes, for which the Euclidean section of the spacetime is regular only if the Euclidean time coordinate is assigned a specific periodicity. To rigorously determine this periodicity, we invoked the CGB theorem, which provides a geometric condition for the regularity of the Euclidean manifold. By focusing on the near-horizon geometry, we demonstrated that the requirement of smoothness in the Euclidean sector imposes a unique periodicity on the Euclidean time coordinate. This, in turn, leads directly to a unique and well-defined expression for the black hole entropy, consistent with the Bekenstein-Hawking area law.\\

In contrast, the situation becomes more subtle in the case of extremal black holes. For these solutions, the Euclidean geometry differs qualitatively from the non-extremal case due to the degeneracy of the horizon and the resulting change in the structure of the near-horizon region. We showed that neither the near-horizon analysis nor the application of the CGB theorem is sufficient to uniquely determine the periodicity of the Euclidean time coordinate in the extremal limit. The absence of a conical singularity in the Euclidean continuation of extremal black holes implies that the periodicity of the Euclidean time remains undetermined. Consequently, the entropy of extremal black holes cannot be uniquely derived using the Euclidean path integral method alone. This ambiguity highlights a fundamental limitation of the semiclassical Euclidean approach when extended to extremal geometries. It suggests that additional input, possibly from quantum gravity or microstate counting methods, is required to resolve the entropy in such cases.\\


\appendix

\bibliography{BibTex}



\end{document}